\documentclass[preprint,aps,a4,amsmath,amssymb,floatfix]{revtex4}

\usepackage{bm}
\usepackage{graphicx}
\newcommand{\vlk}{$V_{\rm low-k}$ }
\newcommand{\vlkn}{$V_{\rm low-k}$}
\newcommand{\be}{\begin{equation}}
\newcommand{\ee}{\end{equation}}

\begin{document} 

\title{Density-dependent effective nucleon-nucleon interaction \\ from chiral three-nucleon 
forces\footnote{Work supported in part by BMBF, GSI and by the DFG
cluster of excellence: Origin and Structure of the Universe.}}

\author{J.\ W.\ Holt, N.\ Kaiser, and W.\ Weise}
\affiliation{Physik Department, Technische Universit\"{a}t M\"{u}nchen,
    D-85747 Garching, Germany}

\begin{abstract}
We derive density-dependent corrections to the in-medium nucleon-nucleon
interaction from the leading-order chiral three-nucleon force. 
To this order there are six distinct one-loop diagrams contributing
to the in-medium nucleon-nucleon scattering $T$-matrix.
Analytic expressions are presented for each of these in both 
isospin-symmetric nuclear matter as well as nuclear 
matter with a small isospin asymmetry. The results are combined with the 
low-momentum nucleon-nucleon potential \vlk to obtain an effective 
density-dependent interaction suitable for nuclear structure calculations. The 
in-medium interaction is decomposed into partial waves up to orbital angular 
momentum $L=2$. Our results should be particularly useful in calculations 
where an exact treatment of the chiral three-nucleon force would otherwise 
be computationally prohibitive.
\end{abstract}

\maketitle

\section{Introduction}
Three-body forces play an essential role in describing the detailed
properties of nuclear many-body
systems. Compelling evidence for this arises from the inability of
non-relativistic nucleon-nucleon (NN) interactions alone to accurately 
reproduce (i) the saturation properties of nuclear 
matter, (ii) the binding energies and spectra
of light nuclei, and (iii) nucleon-deuteron scattering differential
cross sections at intermediate energies. These systems are simple enough that 
the uncertainties associated with the solution of the corresponding many-body 
equation are well controlled, and therefore deviations from experimental data 
are interpreted as deficiencies in the underlying nuclear force model.

In the case of nuclear matter, a variety of methods such as the 
Brueckner-Bethe-Goldstone hole-line expansion \cite{day67,baldo98}, 
variational methods \cite{day78,lagaris}, and the summation of 
particle-particle hole-hole ring diagrams \cite{yang86,song87} all produce 
similar results for the energy per particle as a function of the 
density. However, these calculations result in a series of saturation points
that lie at either too high a density or too low a binding energy
compared to the empirical values of $\rho_0 = 0.16$ fm$^{-3}$ and $E/A = -16$
MeV. Numerous solutions to this problem have been
proposed, such as including relativistic effects \cite{brockmann,dalen}, 
three-nucleon forces \cite{fabroccini,achim}, or medium-modified meson masses 
\cite{rapp,siu09}. The analogous problem in light nuclei is that calculated
binding energies come out $\sim 10-15$\% smaller
than experimental values 
when two-body interactions alone are used \cite{friar,nogga00,carlson87}.
By including various models of the three-nucleon force, several groups have 
shown that not only the nuclear binding energy problem can be
remedied but the low-lying spectra of light nuclei can also be 
greatly improved \cite{nogga00,pieper02,navratil}. Finally, the differential 
cross section for elastic nucleon-deuteron scattering at intermediate energies
($E^{\rm lab}_N = 100-300$ MeV) exhibits a minimum at backward angles that is
larger by nearly 
30\% compared to Faddeev calculations using NN interactions alone 
\cite{sakamoto,sakai}. This discrepancy is well accounted for when effects from 
three-body forces \cite{witala} or (equivalently) explicit $\Delta$ isobar degrees of 
freedom \cite{sakamoto, nemoto} are included.

The preceding examples illustrate the limitations in obtaining precise
agreement with various nuclear observables when using only two-body interactions
that are fit to free-space NN scattering data. Although there has been much
effort devoted to the construction of realistic models of the three-nucleon
force, it is technically very challenging to include them
in large-scale calculations of medium and heavy nuclei. An alternative and 
simpler approach is to employ instead density-dependent two-body interactions
that reflect the underlying three-nucleon forces. Therefore,
the purpose of the present work is to construct a density-dependent in-medium
NN interaction that is generated at one-loop order by the genuine 
three-nucleon force. We have found in ref.\ \cite{holt09} that such density-dependent 
corrections to the NN interaction strongly suppress the $^{14}$C ground state to $^{14}$N 
ground state Gamow-Teller transition and in this way helps to explain the anomalously 
long lifetime of $^{14}$C.
We work within the framework of chiral effective field
theory and take into account the leading-order three-nucleon forces arising in
that systematic approach. We consider first the density-dependent terms in
isospin symmetric nuclear matter and then continue with the leading (linear) 
effects due to an isospin asymmetry, which are likely 
to be relevant for heavy nuclei. The paper is organized as follows. 
In section \ref{nieft} we briefly describe the chiral effective field theory 
approach to nuclear interactions, together with the explicit form of
the leading-order three-nucleon force. In section \ref{imni} 
we evaluate the one-loop corrections to NN scattering in the nuclear medium
generated by this chiral three-nucleon force. In section \ref{nr} we present
numerical results for the in-medium NN interaction in partial waves up to 
orbital angular momentum $L=2$. The paper ends with a summary and conclusions.

\section{Nuclear interactions from effective field theory}
\label{nieft}
\subsection{Chiral effective interactions}

Chiral effective field theory provides a useful tool for describing 
a wide range of low-energy hadronic phenomena within a single consistent framework. 
The theory exploits the natural separation of energy scales that results from
the spontaneous (and explicit) breaking of chiral symmetry in QCD, which gives 
rise to light pseudoscalar Goldstone bosons (the pion triplet in the two flavor
case). These, together with nucleons, comprise the low energy degrees of 
freedom of the theory. Long-range effects from one- and two-pion exchange between 
nucleons are treated explicitly, while the short-distance dynamics due to 
heavy mesons and baryon resonances (with the possible exception of the 
$\Delta$ isobar) are integrated out and their effects are encoded in nucleon 
contact terms. Presently the computation of the chiral NN potential has been 
carried out to next-to-next-to-next-to-leading order (N$^3$LO) in the small 
momentum expansion \cite{kaiser01,kaiser02,entem,entem2}. At this order it is 
possible to achieve an agreement with empirical NN scattering phase shifts that
is comparable to previous high-precision NN potentials 
\cite{cdbonn,nijmegen,argonne}. By adjusting the 29 parameters (mostly 
low-energy constants) that occur at this order, the 1999 database for $np$ and
$pp$ elastic scattering up to $E_{\rm lab} = 290$ MeV can be fit with a 
$\chi^2$/dof of 1.1 for $np$ scattering and 1.5 for $pp$ scattering. 
Furthermore, the experimental deuteron binding energy, charge radius, and electric 
quadrupole moment are very well reproduced by the chiral N$^3$LO potential \cite{entem2}.
When applied to two- and few-nucleon problems, these chiral potentials are 
regulated by exponential functions \cite{entem, epelbaum2} with cutoffs ranging
from 500 to 700 MeV in order to eliminate high-momentum components. In addition, in-medium
chiral perturbation theory which emphasizes the role of explicit two-pion-exchange dynamics
in nuclear matter has been developed in ref.\ \cite{fritsch}.

\begin{figure}
\begin{center}
\includegraphics[height=5cm]{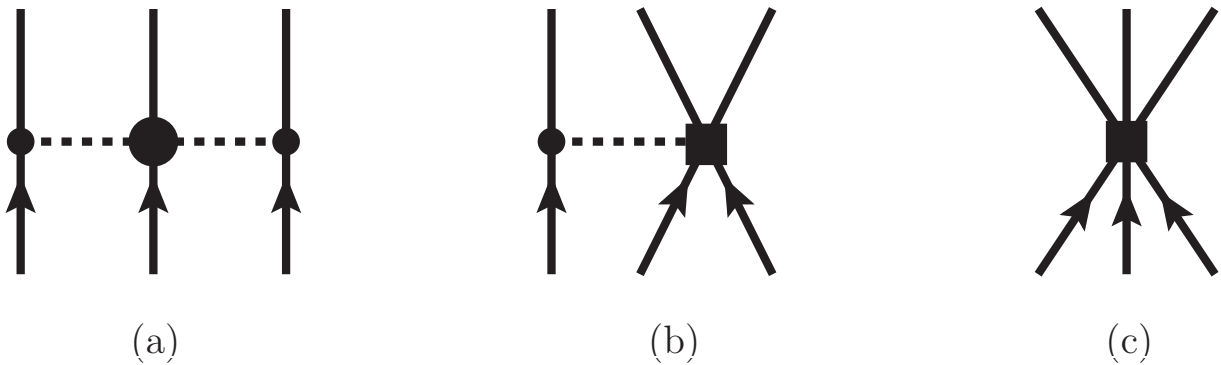}
\end{center}
\vspace{-.5cm}
\caption{The leading-order contributions to the chiral three-nucleon force: (a)
the long-range $2\pi$-exchange force $V_{3N}^{(2\pi)}$, (b) the medium-range 
$1\pi$-exchange force $V_{3N}^{(1\pi)}$, and (c) the short-range contact interaction
$V_{3N}^{(\rm ct)}$.}
\label{tnff}
\end{figure}

Three-nucleon forces arise first at third-order in the chiral power counting
\cite{epelbaumR}. Three components of different range, shown by diagrams (a),(b), and
(c) in Fig.\ \ref{tnff}, occur at this order and have the following analytic structure
\begin{equation}
V_{3N}^{(2\pi)} = \sum_{i\neq j\neq k} \frac{g_A^2}{8f_\pi^4} 
\frac{\vec{\sigma}_i \cdot \vec{q}_i \, \vec{\sigma}_j \cdot
\vec{q}_j}{(\vec{q_i}^2 + m_\pi^2)(\vec{q_j}^2+m_\pi^2)}
F_{ijk}^{\alpha \beta}\tau_i^\alpha \tau_j^\beta,
\label{3n1}
\end{equation}
\begin{equation}
V_{3N}^{(1\pi)} = -\sum_{i\neq j\neq k}
\frac{g_A c_D}{8f_\pi^4 \Lambda_\chi} \frac{\vec{\sigma}_j \cdot \vec{q}_j}{\vec{q_j}^2+m_\pi^2}
\vec{\sigma}_i \cdot
\vec{q}_j \, {\vec \tau}_i \cdot {\vec \tau}_j ,
\label{3n2}
\end{equation}
\begin{equation}
V_{3N}^{(\rm ct)} = \sum_{i\neq j\neq k} \frac{c_E}{2f_\pi^4 \Lambda_\chi}
{\vec \tau}_i \cdot {\vec \tau}_j,
\label{3n3}
\end{equation}
where $\vec{q}_i=\vec{p_i}^\prime -\vec{p}_i$ is the difference between the 
final and initial momentum of nucleon $i$ and 
\begin{equation}
F_{ijk}^{\alpha \beta} = \delta^{\alpha \beta}\left (-4c_1m_\pi^2
 + 2c_3 \vec{q}_i \cdot \vec{q}_j \right ) + 
c_4 \epsilon^{\alpha \beta \gamma} \tau_k^\gamma \vec{\sigma}_k
\cdot \left ( \vec{q}_i \times \vec{q}_j \right ).
\label{3n4}
\end{equation}
Together with $g_A=1.29$ and $f_\pi = 92.4$ MeV, the parameters of $V_{3N}^{(2\pi)}$, namely $c_1 =-0.81\,$GeV$^{-1}$,
$c_3=-3.2\,$GeV$^{-1}$, and $c_4 =5.4\,$GeV$^{-1}$, are well determined from
fits to low-energy NN phase shifts and mixing angles \cite{entem}. Restricting
the two large coefficients 
$c_{3,4}$ to their dominant $\Delta(1232)$-resonance contributions, one
reproduces the celebrated three-nucleon force of Fujita and Miyazawa 
\cite{fujita}. The medium-range ($V_{3N}^{(1\pi)}$) and short-range ($V_{3N}^{(\rm ct)}$) 
components are proportional to two new 
low-energy constants $c_D$ and $c_E$, respectively. These constants can be 
fixed by fitting 
properties of few-nucleon systems, such as the triton binding energy 
together with the $nd$ doublet scattering length \cite{epelbaum02},
the $^4$He binding energy \cite{nogga06}, or binding energies and spectra
of light nuclei \cite{navratil}. At next order (in the chiral expansion) there 
are many additional one-loop diagrams contributing to the chiral 
three-nucleon force \cite{epelbaumR}, but no new low-energy constants 
appear. The explicit construction of all these terms is currently 
in progress.

\subsection{Low-momentum nuclear interaction}
Recently there has been much interest in understanding the scale-dependence
of NN interactions from the point of view of 
the renormalization group \cite{bogner02,bogner03}. In most applications of 
chiral effective field theory to few-nucleon systems, exponential regulator functions 
with cutoffs between 500 and 700 MeV are used to eliminate
the high-momentum components from the NN interaction. In fact, all 
realistic NN potentials are fit to $pp$ and $pn$ scattering data below
a laboratory energy of 350 MeV and therefore are constrained experimentally 
only up to a momentum of $p_{\rm max} \simeq 400$ MeV $= 2.1$ 
fm$^{-1}$. Bogner 
and collaborators \cite{bogner02,bogner03} have shown how to evolve any
bare NN interaction down to a low-momentum scale via 
renormalization group techniques. Such low-momentum interactions \vlk are
phase-shift equivalent to the underlying bare interaction at energies below the
specified cutoff scale $\Lambda_{\rm low-k}$. Moreover, when the decimation
scale is reduced to $p_{\rm max}$, all realistic NN interactions merge
to a nearly universal potential, thereby removing the model dependence related to 
the high momentum components. The method for constructing such low-momentum 
interactions is as follows.

One begins with the half-on-shell $T$-matrix 
for free NN scattering in a given partial wave,
\begin{equation}
T(p',p) = V_{NN}(p',p) + \frac{2}{\pi}{\cal P} \int _0 ^{\infty} 
\frac{V_{NN}(p',q)T(q,p)} {p^2-q^2} q^2 dq,
\end{equation}
and introduces a low-momentum half-on-shell $T$-matrix at the scale $\Lambda_{\rm low-k}$
\begin{equation}
T_{\rm low-k }(p',p) = V_{\rm low-k}(p',p) + \frac{2}{\pi}{\cal P} 
\int _0 ^{\Lambda_{\rm low-k}} \frac{V_{\rm low-k }(p',q) T_{\rm low-k} 
(q,p)}{p^2-q^2} q^2 dq,
\label{vlke}
\end{equation}
where $\cal P$ denotes the principal value. Since one has to preserve
low-energy observables, e.g. scattering phase shifts defined through
$\tan \delta(p) = -p\, T(p,p)$, one requires that
\begin{equation}
T_{\rm low-k}(p',p) = T(p',p) \, , \hspace{.1in} {\rm for} \hspace{.2in} p',p 
< \Lambda_{\rm low-k}.
\label{pse}
\end{equation}
Eqs. (\ref{vlke}) and (\ref{pse}) together define the low-momentum NN interaction 
\vlk (for a review see ref.\ \cite{bogner03}). Under 
the scale decimation procedure, all high-precision NN potentials
flow, as \mbox{$\Lambda_{\rm low-k} \rightarrow 2.1$ $\rm fm^{-1}$}, to a 
nearly unique low-momentum potential $V_{\rm low-k}$. In the present study
we employ the (bare) chiral N$^3$LO interaction and evolve it down to a 
resolution scale $\Lambda_{\rm low-k}=2.1$ fm$^{-1}$.

As demonstrated in ref.\ \cite{achim} the addition of three-nucleon 
forces is essential for obtaining reasonable
saturation properties of nuclear matter when using the universal 
low-momentum NN potential $V_{\rm low-k}$ in Hartree-Fock 
calculations. The construction of decimated low-momentum three-body forces
consistent with the two-body decimation for \vlk remains a challenge.
The difficulty has so far been addressed pragmatically by exploiting 
that low-energy nuclear observables must be scale-independent. Following
this reasoning the low-energy constants associated with the
one-pion exchange component $c_D(\Lambda_{\rm low-k})$ and the short-range 
contact term $c_E(\Lambda_{\rm low-k})$ 
of the chiral three-nucleon interaction have been fitted to
experimental binding energies of three- and four-nucleon systems 
($^3$H, $^3$He, and $^4$He) at a given (variable) low-momentum scale 
$\Lambda_{\rm low-k}$. Since we investigate the density-dependent two-nucleon
interaction at the scale $\Lambda_{\rm low-k} = 2.1$ fm$^{-1}$, we use the 
corresponding values for $c_D$ and $c_E$ obtained in ref.\ \cite{nogga}:
\begin{equation}
c_D(2.1 \hspace{.05in}{\rm fm}^{-1})= -2.062 \,, \qquad c_E(2.1 \hspace{.05in}{\rm fm}^{-1})= -0.625 \,,
\end{equation}
together with $\Lambda_{\chi} = 0.7$ GeV.
In addition to the low-momentum decimation techniques described above, there 
exist several other means to construct effective nucleon-nucleon interactions 
suitable for perturbative many-body calculations, such as the similarity 
renormalization group transformations \cite{bogner07} and the unitary correlation
operator method (UCOM) \cite{feldmeier,neff}. The resulting effective two-body 
interactions are all quantitatively similar, and the inclusion of three-nucleon forces
within these different frameworks is currently in progress.

\section{In-medium nucleon-nucleon interaction}
\label{imni}
\subsection{Density-dependent terms in isospin-symmetric nuclear matter}
\label{simni}

In this section we derive from the leading-order chiral three-nucleon 
interaction, eqs.\ (\ref{3n1}--\ref{3n4}), an effective density-dependent in-medium NN interaction. 
We are considering the on-shell scattering of two nucleons in isospin-symmetric 
(spin-saturated) nuclear
matter of density $\rho=2k_f^3/3\pi^2$ in the center-of-mass 
frame, $N_1(\vec p\,)+ N_2(-\vec p\,) \to N_1(\vec 
p+\vec q\,) + N_2(-\vec p-\vec q\,)$. The kinematics is such that the total momentum of the
two-nucleon system is zero in the nuclear matter rest frame before and after
the scattering. The magnitude of the in- and out-going nucleon momenta is $p
= |\vec p\,| = |\vec p+\vec q\,|$, and $q= |\vec q\,|$ is the magnitude of 
the momentum transfer. The on-shell interaction in momentum-space has the
following (general) form
\begin{eqnarray}
V(\vec p,\vec q) &=& V_C + \vec \tau_1 \cdot \vec \tau_2\, W_C + 
\left [V_S + \vec \tau_1 \cdot \vec \tau_2 \, W_S \right ] \vec \sigma_1 \cdot 
\vec \sigma_2 + \left [ V_T + \vec \tau_1 \cdot \vec \tau_2 \, W_T \right ] 
\vec \sigma_1 \cdot \vec q \, \vec \sigma_2 \cdot \vec q \nonumber \\
&+& \left [ V_{SO} + \vec \tau_1 \cdot \vec \tau_2 \, W_{SO} \right ] \,
i (\vec \sigma_1 + \vec \sigma_2 ) \cdot (\vec q \times \vec p) \nonumber \\
&+& \left [ V_{Q} + \vec \tau_1 \cdot \vec \tau_2 \, W_{Q} \right ] \,
\vec \sigma_1 \cdot (\vec q \times \vec p)\, \vec \sigma_2 \cdot (\vec q \times
\vec p).
\end{eqnarray}
The subscripts refer to the central (C), spin-spin (S), tensor (T), 
spin-orbit (SO), and quadratic spin orbit (Q) components of the NN 
interaction, each of which occurs in an isoscalar (V) and an isovector (W) version.
For the purpose of comparison with the density-dependent terms that follow, we 
reproduce the expression for the (bare) $1\pi$-exchange:
\begin{equation} V_{NN}^{(1\pi)} = - {g_A^2 M_N \over 16 \pi f_\pi^2} \vec \tau_1 \cdot  
\vec \tau_2 \,  {\vec \sigma_1 \cdot \vec q \,\vec \sigma_2 \cdot \vec q\over  
m_\pi^2  + q^2}\,. 
\label{ope}
\end{equation}   
Here $\vec \sigma_{1,2}$ and $\vec \tau_{1,2}$ are the
usual spin and isospin operators of the two nucleons. Note that we have included an additional factor
of $M_N/4\pi$ in $V_{NN}$ in order to be consistent with the conventions commonly chosen for \vlkn.

\begin{figure}
\begin{center}
\includegraphics[scale=1.1,clip]{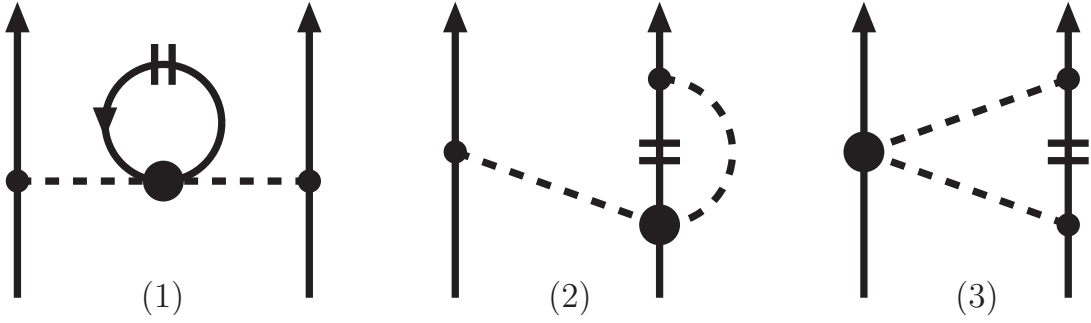}
\end{center}
\vspace{-.5cm}
\caption{In-medium NN interaction generated by the two-pion exchange component
($\sim c_{1,3,4}$) of the chiral three-nucleon interaction. The short 
double-line symbolizes the filled Fermi sea of nucleons, i.e. the medium
insertion $-2\pi \delta(k_0)\, \theta(k_f-|\vec k\,|)$ in the in-medium
nucleon propagator. Reflected diagrams are not shown.}  
\label{mfig1}
\end{figure}

We start with those contributions to the in-medium NN-interaction $V_{NN}^{\rm 
med}$ that are generated by the $2\pi$-exchange component of the chiral
three-nucleon force. The three different topologies for non-vanishing
one-loop diagrams are shown in Fig.\ \ref{mfig1}. The short double-line on 
a nucleon propagator symbolizes the filled Fermi sea of nucleons, which introduces
the medium insertion $-2\pi \delta(k_0)\,
\theta(k_f- |\vec k\,|)$ in the in-medium nucleon propagator. In effect, the
medium insertion sums up hole propagation and the absence of particle
propagation below the Fermi surface $|\vec k\,|<k_f$. The left diagram in 
Fig.\ \ref{mfig1} represents a $1\pi$-exchange with a Pauli blocked in-medium 
pion self-energy and the corresponding contribution to $V_{NN}^{\rm  med}$ reads:
\begin{equation} V_{NN}^{\rm med,1}= {g_A^2 M_N \rho \over 8 \pi f_\pi^4}\,\vec \tau_1  
\cdot \vec \tau_2 \,{\vec \sigma_1 \cdot\vec q \,\vec \sigma_2 \cdot \vec q 
\over (m_\pi^2 + q^2)^2}\,(2c_1 m_\pi^2 +c_3 q^2)\,.
\label{med1}
\end{equation}
Since $c_{1,3}<0$, this term corresponds to an
enhancement of the bare $1\pi$-exchange. It can be interpreted in terms of the 
reduced pion decay constant, $f_{\pi,s}^{*2}=f_\pi^2+2c_3\rho$, that replaces 
$f_\pi^2$ in the denominator of eq.\ (\ref{ope}) and which is associated with 
the space components of the axial current in the nuclear medium. The second
diagram in Fig.\ \ref{mfig1} includes vertex corrections 
to the $1\pi$-exchange caused by Pauli blocking in the nuclear medium. The 
corresponding contribution to the in-medium NN-interaction has the form:
\begin{eqnarray} V_{NN}^{\rm med,2}&=& {g_A^2 M_N \over 32\pi^3 f_\pi^4}\vec \tau_1  
\cdot \vec \tau_2 \,  {\vec \sigma_1 \cdot \vec q \,\vec \sigma_2 \cdot \vec q 
\over m_\pi^2 + q^2}\, \bigg\{-4c_1 m_\pi^2 \Big[\Gamma_0(p)+\Gamma_1(p) \Big]
\nonumber \\ && - (c_3+c_4) \Big[q^2 \Big(\Gamma_0(p)+2\Gamma_1(p)+
\Gamma_3(p)\Big)+ 4\Gamma_2(p)\Big] + 4c_4 \bigg[ {2k_f^3 \over
  3}-m_\pi^2\Gamma_0(p)\bigg] \bigg\}\,.
\label{med2}
\end{eqnarray}
Here, we have introduced the $k_f$-dependent functions $\Gamma_j(p)$ which result from Fermi 
sphere integrals, $\int_{|\vec{k}|\leq k_f} d^3k$, over a static pion-propagator $[m_\pi^2+(\vec k+\vec p\,
)^2]^{-1}$:
\begin{equation} \Gamma_0(p) = k_f - m_\pi \bigg[ \arctan{k_f+p \over m_\pi} +
\arctan{k_f-p \over m_\pi}\bigg] + {m_\pi^2 +k_f^2 -p^2 \over 4p}\ln {m_\pi^2  
+(k_f+p)^2 \over m_\pi^2+(k_f-p)^2} \,, \end{equation}   
\begin{equation} \Gamma_1(p) = {k_f \over 4p^2}  (m_\pi^2+k_f^2+p^2) -\Gamma_0(p)
-{1\over 16 p^3} \Big[m_\pi^2 +(k_f+p)^2\Big]\Big[ m_\pi^2+(k_f-p)^2\Big]\ln 
{m_\pi^2   +(k_f+p)^2 \over m_\pi^2+(k_f-p)^2} \,, \end{equation} 
 \begin{equation} \Gamma_2(p)= {k_f^3 \over 9}+{1\over6} (k_f^2-m_\pi^2-p^2)
\Gamma_0(p)+{1\over6} (m_\pi^2+k_f^2-p^2)\Gamma_1(p)\,, \end{equation}
\begin{equation} \Gamma_3(p)= {k_f^3 \over 3p^2}-{m_\pi^2+k_f^2+p^2 \over  2p^2}
\Gamma_0(p)-{m_\pi^2+k_f^2+3p^2 \over 2p^2}\Gamma_1(p)\,. \end{equation}
By analyzing the momentum and density dependent factor in eq.\ (\ref{med2}) relative to
$V_{NN}^{(1\pi)}$, one finds that this contribution corresponds to a reduction
of the $1\pi$-exchange in the nuclear medium. Approximately, this feature can
be interpreted in terms of a reduced nucleon axial-vector constant
$g_A^*$.

\begin{figure}
\begin{center}
\includegraphics[scale=1.1,clip]{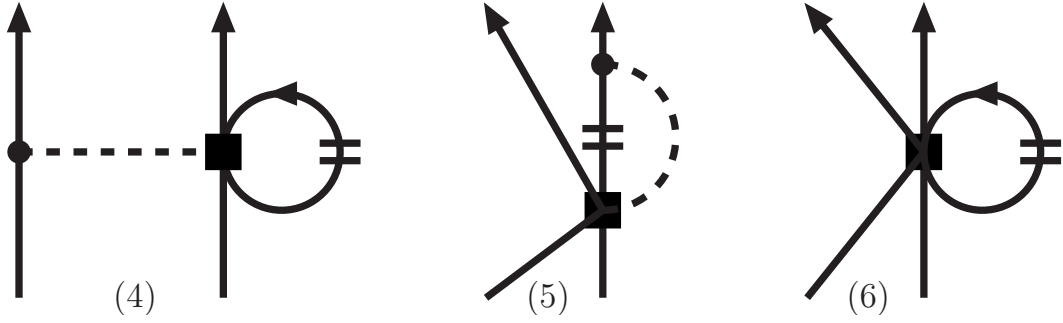}
\end{center}
\vspace{-.5cm}
\caption{In-medium NN interaction generated by the one-pion exchange ($\sim 
c_D$) and short-range component ($\sim c_E$) of the chiral three-nucleon
interaction.} 
\label{mfig2}
\end{figure}

The right diagram in Fig.\ \ref{mfig1} represents Pauli blocking effects on chiral 
$2\pi$-exchange. Evaluating it together with the reflected diagram one finds
the following contribution to the in-medium NN-interaction:    
\begin{eqnarray} V_{NN}^{\rm med,3} &=& {g_A^2 M_N \over 64 \pi^3 f_\pi^4}\bigg\{  -12 c_1 
m_\pi^2 \Big[2\Gamma_0(p)- (2m_\pi^2+q^2) G_0(p,q)\Big] \nonumber \\ && -c_3
\Big[8 k_f^3-12(2m_\pi^2+q^2) \Gamma_0(p) -6q^2\Gamma_1(p)+3(2m_\pi^2+q^2)^2 G_0(p,q)
\Big]\nonumber \\&& +4c_4\,\vec \tau_1 \cdot  \vec \tau_2\, (\vec \sigma_1
\cdot \vec \sigma_2\, q^2 - \vec \sigma_1 \cdot  \vec q \,\vec\sigma_2\cdot
\vec q\,) G_2(p,q) \nonumber \\ && -(3c_3+c_4\vec \tau_1 \cdot \vec \tau_2 )\,
i ( \vec \sigma_1 +\vec \sigma_2 )\cdot(\vec q \times \vec p\,)\Big[2\Gamma_0(p)
+2\Gamma_1(p) - (2m_\pi^2+q^2)\nonumber \\ &&\times \Big(G_0(p,q)+2 G_1(p,q)
\Big) \Big] -12 c_1 m_\pi^2\, i ( \vec \sigma_1 +\vec \sigma_2 ) \cdot(\vec q 
\times \vec p\,)  \Big[G_0(p,q)+2 G_1(p,q)\Big] \nonumber \\ && + 4c_4\, 
\vec\tau_1 \cdot \vec \tau_2 \,\vec \sigma_1 \cdot (\vec q \times \vec p\,)\,
\vec\sigma_2 \cdot( \vec q \times \vec p \,) \Big[G_0(p,q) +
4G_1(p,q)+4G_3(p,q) \Big]\bigg\}\,. 
\label{med3}
\end{eqnarray}
One observes that in comparison to the analogous $2\pi$-exchange interaction
in vacuum (see section 4.2 in ref.\ \cite{nnpap}) the Pauli blocking in the 
nuclear medium  has generated additional spin-orbit terms, $i(\vec\sigma_1 
+\vec \sigma_2 
)\cdot(\vec q \times \vec p\,)$, and quadratic spin-orbit terms, $\vec \sigma_1 
\cdot (\vec q \times \vec p\,)\,\vec\sigma_2 \cdot( \vec q \times \vec p \,)$, 
written in the last three lines of eq.\ (\ref{med3}). The density dependent spin-orbit
terms (scaling with $c_{3,4}$) in the in-medium NN-interaction $V_{NN}^{\rm  med,3}$ 
demonstrate clearly and explicitly the mechanism of three-body induced
spin-orbit forces proposed long ago by Fujita and Miyazawa \cite{fujita}.  
The functions $G_j(p,q)$ appearing in eq.\ (\ref{med3}) result from Fermi 
sphere
integrals over the product of two  different pion-propagators. Performing the
angular integrations analytically one arrives at:
\begin{equation} G_{0,*,**}(p,q) = {2\over q} \int_0^{k_f}\!\! dk\,  {\{k,k^3,k^5\} 
\over \sqrt{A(p)+q^2 k^2} } \ln { q\, k+\sqrt{A(p)+q^2 k^2}\over \sqrt{A(p)}}\,,
\end{equation} 
with the abbreviation $A(p)= [m_\pi^2 +(k+p)^2][ m_\pi^2+(k-p)^2]$. The other
functions with $j=1,2,3$ are obtained by solving a system of linear equations:
\begin{eqnarray} G_1(p,q)&=& {\Gamma_0(p)-(m_\pi^2+p^2)G_0(p,q) -G_*(p,q) \over 
4p^2-q^2} \,,\\ G_{1*}(p,q)&=&  {3\Gamma_2(p)+p^2\Gamma_3(p)-(m_\pi^2+p^2)G_*(p,q)
  -G_{**}(p,q) \over  4p^2-q^2} \,,\\G_2(p,q)&=&(m_\pi^2+p^2)G_1(p,q)+G_*(p,q)+
G_{1*}(p,q)\,,\\ G_3(p,q)&=& {{1\over 2}\Gamma_1(p)-2(m_\pi^2+p^2)G_1(p,q) 
-2G_{1*}(p,q) -G_{*}(p,q) \over 4p^2-q^2} \,.\end{eqnarray} 
In this chain of equations the functions indexed with an asterisk play 
only an auxiliary role for the construction of $G_{1,2,3}(p,q)$. We note that all 
functions $G_j(p,q)$ are non-singular at $q=2p$ (corresponding to scattering
in backward direction). For notational simplicity, the $k_f$-dependence of $\Gamma_j(p)$
and $G_j(p,q)$ has been suppressed.

Next, we come to the $1\pi$-exchange component of the chiral three-nucleon
interaction proportional to the parameter $c_D/\Lambda_{\chi}$, where $c_D \simeq -2$
for a scale of $\Lambda_{\chi} =0.7\,$GeV \cite{nogga}. The filled black square in
the first and second diagram of Fig.\ \ref{mfig2} symbolizes the corresponding 
two-nucleon one-pion contact interaction. By closing a nucleon line at the
contact vertex, one obtains a vertex correction (linear in the density $\rho$) 
to the  $1\pi$-exchange:
\begin{equation} V_{NN}^{\rm med,4} = -{g_A M_N c_D \rho \over 32 \pi f_\pi^4\Lambda_{\chi}}
\,\vec \tau_1 \cdot \vec \tau_2 \,{\vec \sigma_1 \cdot\vec q \,\vec \sigma_2 
\cdot \vec q  \over  m_\pi^2 + q^2} \,.\end{equation}
Since $c_D$ is negative, this term reduces again the bare $1\pi$-exchange,
roughly by about $16\%$ at normal nuclear matter density $\rho_0=0.16\,
$fm$^{-3}$. The second diagram in Fig.\ \ref{mfig2} includes Pauli blocked (pionic) 
vertex corrections to the short-range NN interaction. The corresponding 
contribution to the density dependent in-medium NN interaction reads:
\begin{eqnarray} V_{NN}^{\rm med,5}&=& {g_A M_N c_D\over 64 \pi^3 f_\pi^4\Lambda_{\chi}}\vec 
\tau_1 \cdot \vec \tau_2 \bigg\{2 \vec \sigma_1 \cdot \vec \sigma_2\,\Gamma_2(p)
 +\bigg[\vec \sigma_1 \cdot \vec \sigma_2 \bigg( 2p^2-{q^2\over 2}\bigg) + \vec 
\sigma_1 \cdot \vec q\,\vec\sigma_2  \cdot \vec q\, 
\bigg(1-{2p^2\over q^2}\bigg) \nonumber \\ &&  -{2\over q^2}\,\vec\sigma_1 \cdot (\vec q \times 
\vec p\,)\,\vec\sigma_2 \cdot(\vec q\times \vec p \,)\bigg]
\Big[\Gamma_0(p)+2\Gamma_1(p)+\Gamma_3(p)\Big] \bigg\}\,, 
\label{med5}
\end{eqnarray}
where we have used an identity for $\vec \sigma_1 \cdot \vec p\,\, \vec\sigma_2 
\cdot\vec p+ \vec \sigma_1 \cdot (\vec p+\vec q\,)\, \vec\sigma_2\cdot (\vec p+
\vec q\,)=[\dots]$. The ellipses stands for the combination of spin operators 
written in the square bracket of eq.\ (\ref{med5}).  

Finally, there is the short-range component of the chiral 3N interaction, 
represented by a three-nucleon contact-vertex proportional to  $c_E/\Lambda_{\chi}$. By
closing one nucleon line (see right diagram in Fig.\ 2) one obtains the
following contribution to the in-medium NN-interaction:
\begin{equation}  V_{NN}^{\rm med,6} =-{3 M_N c_E \rho \over 8 \pi f_\pi^4\Lambda_\chi} \,,
\end{equation}
which simply grows linearly in density $\rho=2k_f^3/3\pi^2$ and is independent of 
spin, isospin and nucleon momenta \footnote{In order to facilitate the 
computation of symmetry factors and spin and isospin traces, we have modeled 
(for that purpose)
the three-nucleon contact-interaction by heavy isoscalar boson exchanges.}.

The above expressions have been derived for on-shell scattering, which greatly 
simplifies the calculation of the density-dependent corrections to the NN
interaction. As discussed in ref.\ \cite{holt09}, a suitable choice for 
extrapolating these expressions off-shell is to make the substitution
$p^2\rightarrow \frac{1}{2}(p^2 + {p^\prime}^2)$. Additionally we include
a regulator function of the form
${\rm exp}\,\left [-(p/\Lambda_{\rm low-k})^4 -
(p^\prime/\Lambda_{\rm low-k})^4\right ]$,
where $\Lambda_{\rm low-k}=2.1$ fm$^{-1}$ is chosen so that the two-body and
three-body force contributions are decimated down to the same scale. This is 
consistent with the approach taken in ref.\ \cite{nogga}.

\subsection{Correction terms in isospin asymmetric nuclear matter}
\label{ianm}

Now we consider the additional modifications $W_{NN}^{\rm med,i}$ to the six in-medium NN
scattering $T$-matrices, $V_{NN}^{\rm med,i}$, resulting from a small isospin asymmetry. The total 
nucleon density made up by protons and neutrons is $\rho = \rho_n+\rho_p$, and the Fermi momentum
$k_f$ is given as before by
the relation $\rho = 2k_f^3/3\pi^2$. As a measure of the isospin asymmetry, we define the 
relative neutron excess
$\delta_{np} = (\rho_n-\rho_p)/\rho$. For a heavy nucleus such as $^{208}$Pb,
the relative neutron excess is of the order $\delta_{np} \simeq 0.2$. 
Re-evaluating the six diagrams in Figs.\ \ref{mfig1} and \ref{mfig2} with the substitution
\begin{eqnarray}
\theta(k_f-|\vec{k}|)&\rightarrow& \frac{1+\tau^3}{2}\, \theta (k_f(1-\delta_{np})^{1/3}-|\vec{k}| )
+ \frac{1-\tau^3}{2}\, \theta (k_f(1+\delta_{np})^{1/3}-|\vec{k}|) \nonumber \\ 
&=&\left(1-\tau^3\delta_{np} \frac{k_f}{3}\frac{\partial}{\partial k_f}\right )\theta(k_f-|\vec{k}|)+\cdots,
\end{eqnarray}
we obtain the following expressions for the corrections (linear in $\delta_{np}$) 
to the density-dependent NN interaction.

\begin{equation}
W_{NN}^{\rm med,1}= 0,
\label{medi1}
\end{equation}

\begin{eqnarray}
W_{NN}^{\rm med,2}&=& {g_A^2 M_N \delta_{np} \over 192 \pi^3 f_\pi^4}
(\tau_1^3+\tau_2^3) \,  {\vec \sigma_1 \cdot \vec q \,\vec \sigma_2 \cdot \vec q 
\over m_\pi^2 + q^2}\, k_f\frac{\partial}{\partial k_f}\bigg\{4c_1 m_\pi^2 
\Big[\Gamma_0(p)+\Gamma_1(p) \Big] + (c_3-c_4)
\nonumber \\ &&  \times \Big[q^2 \Big(\Gamma_0(p)+2\Gamma_1(p)+
\Gamma_3(p)\Big)+ 4\Gamma_2(p)\Big] + 4c_4 \bigg[ {2k_f^3 \over
  3}-m_\pi^2\Gamma_0(p)\bigg] \bigg\}\,,
\label{medi2}
\end{eqnarray}

\begin{eqnarray}
W_{NN}^{\rm med,3} &=& {g_A^2 M_N \delta_{np} \over 384 \pi^3 f_\pi^4}
(\tau_1^3+\tau_2^3) k_f\frac{\partial}{\partial k_f}\bigg\{  -4 c_1 
m_\pi^2 \Big[2\Gamma_0(p)- (2m_\pi^2+q^2) G_0(p,q)\Big] \nonumber \\ && -c_3
\Big[\frac{8 k_f^3}{3}-4(2m_\pi^2+q^2) \Gamma_0(p) -2q^2\Gamma_1(p)+
(2m_\pi^2+q^2)^2 G_0(p,q)
\Big]\nonumber \\&& +4c_4\, (\vec \sigma_1
\cdot \vec \sigma_2\, q^2 - \vec \sigma_1 \cdot  \vec q \,\vec\sigma_2\cdot
\vec q\,) G_2(p,q) \nonumber \\ && -(c_3+c_4)\,
i ( \vec \sigma_1 +\vec \sigma_2 )\cdot(\vec q \times \vec p\,)\Big[2\Gamma_0(p)
+2\Gamma_1(p) - (2m_\pi^2+q^2)\nonumber \\ &&\times \Big(G_0(p,q)+2 G_1(p,q)
\Big) \Big] -4 c_1 m_\pi^2\, i ( \vec \sigma_1 +\vec \sigma_2 ) \cdot(\vec q 
\times \vec p\,)  \Big[G_0(p,q)+2 G_1(p,q)\Big] \nonumber \\ && + 4c_4\, 
\vec \sigma_1 \cdot (\vec q \times \vec p\,)\,
\vec\sigma_2 \cdot( \vec q \times \vec p \,) \Big[G_0(p,q) +
4G_1(p,q)+4G_3(p,q) \Big]\bigg\}\,,
\label{medi3}
\end{eqnarray}

\begin{equation}
W_{NN}^{\rm med,4} = {g_A M_N c_D (\rho_p -\rho_n) \over 64 \pi f_\pi^4\Lambda_{\chi}}
\,( \tau_1^3 + \tau_2^3) \,{\vec \sigma_1 \cdot\vec q \,\vec \sigma_2 
\cdot \vec q  \over  m_\pi^2 + q^2} \,,
\label{medi4}
\end{equation}

\begin{eqnarray}
W_{NN}^{\rm med,5}&=& -{g_A M_N c_D\delta_{np} \over 384 \pi^3 f_\pi^4\Lambda_{\chi}}( 
\tau_1^3 + \tau_2^3) k_f \frac{\partial}{\partial k_f}\bigg\{2 \vec \sigma_1 
\cdot \vec \sigma_2\,\Gamma_2(p) +\bigg[\vec \sigma_1 \cdot \vec \sigma_2 
\bigg( 2p^2-{q^2\over 2}\bigg) +  \vec 
\sigma_1 \cdot \vec q\,\vec\sigma_2  \cdot \vec q\, 
\nonumber \\ && \times \bigg(1-{2p^2\over q^2}\bigg)   -{2\over q^2}\,\vec\sigma_1 \cdot (\vec q \times 
\vec p\,)\,\vec\sigma_2 \cdot(\vec q\times \vec p \,)\bigg]
 \Big[\Gamma_0(p)+2\Gamma_1(p)+\Gamma_3(p)\Big] \bigg\}\,, \hspace{.2in} {\rm and}
\label{medi5}
\end{eqnarray}

\begin{equation}
W_{NN}^{\rm med,6} = {3 M_N c_E (\rho_p -\rho_n) \over 16 \pi f_\pi^4\Lambda_\chi}(\tau_1^3+\tau_2^3) \, .
\label{medi6}
\end{equation}

\noindent For completeness we also list the derivatives with respect to $k_f$ of the loop functions 
$\Gamma_j(p)$ and $G_j(p,q)$ encountered in the previous section:
\be
\frac{\partial \Gamma_0(p)}{\partial k_f} = \frac{k_f}{2p}\,{\rm ln}\, 
\frac{m_\pi^2+(k_f+p)^2}{m_\pi^2+(k_f-p)^2},
\ee
\be
\frac{\partial \Gamma_1(p)}{\partial k_f} = \frac{k_f^2}{p^2} - 
\frac{m_\pi^2+k_f^2+p^2}{2p^2}\frac{\partial \Gamma_0(p)}{\partial k_f},
\ee
\be
\frac{\partial \Gamma_2(p)}{\partial k_f} = \frac{k_f^2}{4p^2} 
(m_\pi^2+k_f^2+p^2) -\frac{1}{8p^2}\left [m_\pi^2 + (k_f+p)^2\right ]
\left [m_\pi^2 + (k_f-p)^2\right ]\frac{\partial \Gamma_0(p)}{\partial k_f},
\ee
\be
\frac{\partial \Gamma_3(p)}{\partial k_f} = \frac{1}{p^2}  
\left[ k_f^2 \frac{\partial \Gamma_0(p)}{\partial k_f} 
-3\frac{\partial \Gamma_2(p)}{\partial k_f}\right ],
\ee
\be
\frac{\partial}{\partial k_f}G_{0,*,**}(p,q) = 
\frac{2\{k_f,k_f^2,k_f^3\}}{q\sqrt{\tilde A(p)+q^2 k_f^2}}\, {\rm ln}
\, \frac{q k_f+\sqrt{\tilde A(p)+q^2 k_f^2}}{\sqrt{\tilde A(p)}},
\ee 
with the abbreviation $\tilde A(p) = \left [m_\pi^2 + (k_f+p)^2\right ]
\left [m_\pi^2 + (k_f-p)^2\right ]$. Aside from the pion self-energy term, which 
receives no modification linear in the isospin asymmetry, each term 
$W_{NN}^{\rm med,i}$ is 
proportional to the factor $(\tau_1^3+\tau_2^3)\,\delta_{np}$. Therefore, the 
proton-proton ($pp$) and neutron-neutron ($nn$) interactions in isospin-asymmetric
nuclear matter receive corrections of opposite sign, while the neutron-proton ($np$)
interaction remains unchanged at linear order in $\delta_{np}$. In general, we expect these 
corrections to be small, though the difference between the in-medium $pp$ and
$nn$ interactions can be important as we discuss in the next section.

\section{Numerical results}
\label{nr}
\subsection{Partial wave projection}
For the purpose of demonstration and for nuclear structure calculations it is 
convenient to evaluate the above in-medium NN interaction
in the partial-wave basis consisting of eigenstates with 
definite $L,S,J$ quantum numbers. We have followed the detailed
description in ref.\ \cite{erkelenz} and obtain with
$U_K = V_K+(4I-3) W_K \,\,, (K=C,S,T,SO,Q)$
the relevant linear combination in a state with total isospin $I=0,1$ the
following projection formulas:

\noindent
a) Singlet matrix element with $S=0$ and $L=J$:
\begin{equation}\langle J0J|V_{NN}|J0J\,\rangle 
={1\over2}\int_{-1}^1 dz
\Big[ U_C-3U_{S}-q^2 U_T+p^4(z^2-1) U_Q \Big] P_J(z)\,.\end{equation}
b) Triplet matrix element with $S=1$ and $L=J$:
\begin{eqnarray} & &\langle J1J| V_{NN}|J1J\,\rangle={1\over2}\int_{-1}^1
dz \Big\{2p^2\Big[ U_{SO}-U_T+p^2z U_Q \Big]\Big(P_{J+1}(z)+P_{J-1}(z)
\Big) \nonumber \\ & & \qquad \qquad+\Big[ U_C+ U_{S}+2p^2(1+z)  U_T-
4p^2z U_{SO}-p^4(3z^2+1) U_Q \Big] P_J(z) \Big\}\,. \end{eqnarray}
c) Triplet matrix elements with $S=1$ and $L=J\pm1$:
\begin{eqnarray} & & \langle J\pm1,1J| V_{NN}|J\pm1,1J\,\rangle={1\over2}
\int_{-1}^1 dz \Big\{2p^2\Big[ U_{SO}\pm{1\over2J+1}\Big( U_T-p^2z
U_Q\Big)\Big] P_{J}(z) \nonumber \\ & & \quad \quad+\Big[ U_C+ U_{S}+p^2
\Big( p^2(1-z^2) U_Q-2z U_{SO} \pm{2\over 2J+1} \Big( p^2 U_Q-
U_T \Big) \Big)  \Big] P_{J\pm1}(z) \Big\} \,.\end{eqnarray}
d) Triplet mixing matrix element with $S=1$, $L'=J-1$ and $L=J+1$:
\begin{eqnarray}  \langle J-1,1J| V_{NN}|J+1,1J\,\rangle&=&{\sqrt{J+1}p^2
\over \sqrt{J}(2J+1)} \int_{-1}^1 dz \Big\{\Big(U_T-p^2 U_Q\Big)
P_{J+1}(z)  \nonumber \\ & & +\Big[\big(2J-z(2J+1)\big) U_T+p^2z U_Q
\Big] P_J(z) \Big\} \,. \end{eqnarray}
Here, $P_J(z)$ are ordinary Legendre polynomials of degree $J$ and the 
momentum transfer $q$ is set to $q = p \sqrt{2(1-z)}$. The resulting partial
wave amplitudes are now functions of the momentum $p$ and $k_f$ (or 
equivalently the nucleon density $\rho$). Note that the off-diagonal mixing 
matrix elements arise exclusively from the tensor
operator  $\vec \sigma_1 \cdot  \vec q \,\vec\sigma_2\cdot \vec q$ and the 
quadratic spin-orbit operator  $\vec \sigma_1 \cdot (\vec q \times \vec p\,)\,
\vec\sigma_2 \cdot( \vec q \times \vec p \,)$.

\subsection{$S$-waves and $S-D$ mixing}

\begin{figure}[t]
\begin{center}
\includegraphics[height=13cm,angle=270]{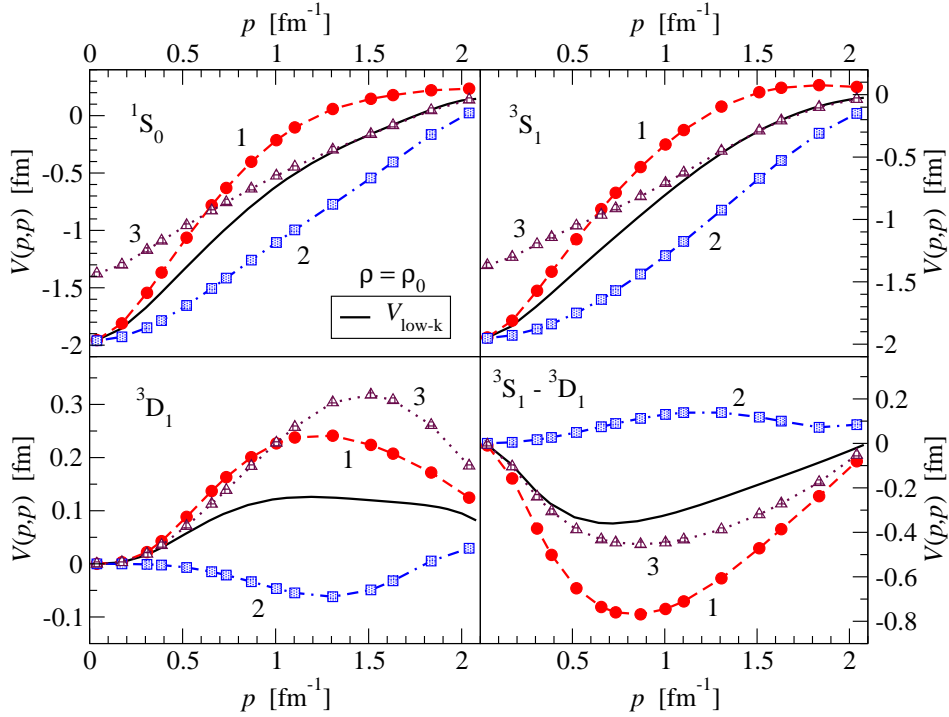}
\end{center}
\vspace{-.5cm}
\caption{Modifications to the $^1$S$_0$, $^3$S$_1$, $^3$D$_1$, and 
$^3$S$_1-^3$D$_1$ partial wave amplitudes of \vlk (shown by the solid
curve) due to the first three density-dependent
contributions $V_{NN}^{\rm med; \, 1,2,3}$ at nuclear matter 
saturation density $\rho_0=0.16$ fm$^{-3}$.}
\label{sd123}
\end{figure}

\begin{figure}[tb]
\begin{center}
\includegraphics[height=13cm,angle=270]{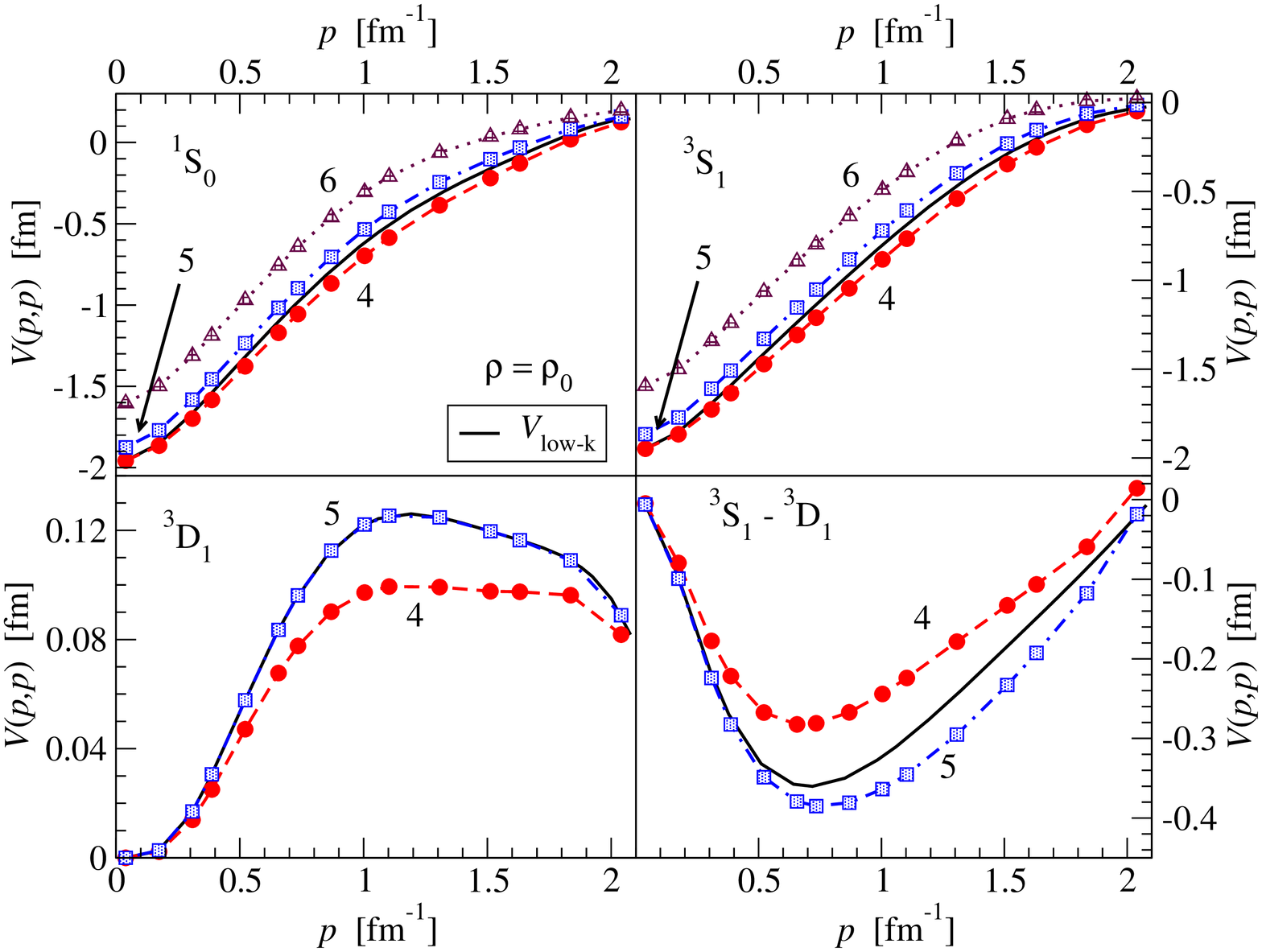}
\end{center}
\vspace{-.5cm}
\caption{Modifications to the $^1$S$_0$, $^3$S$_1$, $^3$D$_1$, and 
$^3$S$_1-^3$D$_1$ partial wave amplitudes of \vlk (shown by the solid 
curve) due to the three density-dependent
contributions $V_{NN}^{\rm med; \, 4,5,6}$
at nuclear matter saturation density $\rho_0=0.16$ fm$^{-3}$.}
\label{sd456}
\end{figure}

In Figs.\ \ref{sd123} and \ref{sd456} we show by the solid lines the (diagonal)
partial-wave amplitudes of \vlk in the channels $^1$S$_0$, $^3$S$_1$, and $^3$D$_1$ as well as 
the $^3$S$_1-^3$D$_1$ mixing matrix element for a cutoff $\Lambda_{\rm low-k} = $ 2.1 fm$^{-1}$.
We also show the modification of the low-momentum interaction due to the six 
different components of the density-dependent NN interaction derived from the 
leading-order chiral three-nucleon force in an isospin symmetric nuclear medium
with density $\rho=\rho_0$. The pion self-energy correction $V_{NN}^{\rm med, 1}$ and the Pauli-blocked 
one-pion exchange vertex correction $V_{NN}^{\rm med, 2}$, which both arise from the long-range 
three-nucleon force, vanish as $q\rightarrow 0$ and therefore give no
modification to the $S$-wave amplitudes 
as $p\rightarrow 0$. Moreover, to a 
large extent they contribute with opposite sign and equal strength. Consequently,
their combined effect on the $S$-wave amplitudes is rather small. 
In comparison, the long-range Pauli-blocked $2\pi$-exchange term $V_{NN}^{\rm med, 3}$ 
contributes strongly to the $S$ waves at large distances (small 
momenta) and has little effect at short distances (large momenta). In fact, 
Pauli-blocked $2\pi$-exchange is the dominant mechanism for suppressing the 
$S$-wave attraction at low momenta.
From Fig.\ \ref{sd456} we observe that the contributions 
$V_{NN}^{\rm med, 4}$ and $V_{NN}^{\rm med, 5}$, resulting from the 
medium-range three-nucleon force proportional to $c_D$, are naturally small in
$S$-waves. Additionally, they enter with opposite sign and therefore
play a very minor role in modifying the $S$-wave interaction. The final
contribution $V_{NN}^{\rm med, 6}$ comes from the three-nucleon contact interaction 
proportional to $c_E$. Since it is independent of the momentum $p$
(aside from the suppression above $p=1.5$ fm$^{-1}$ due to the regulator function), it affects only
the two $S$-waves. 
In combination with $V_{NN}^{\rm med, 3}$ it plays a major role in 
reducing the $S$-wave NN attraction in the nuclear medium.

\begin{figure}
\begin{center}
\includegraphics[height=12cm,angle=270]{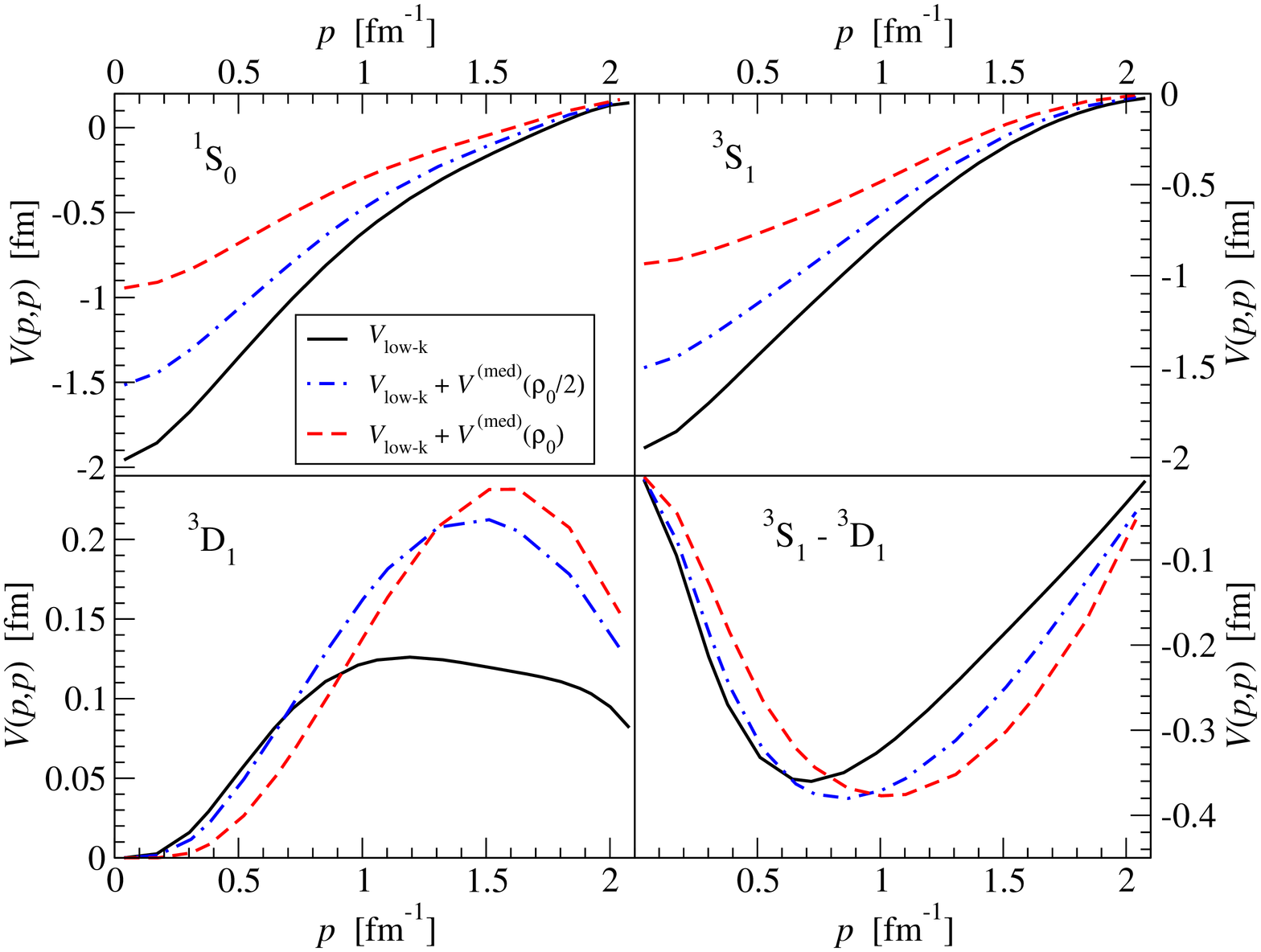}
\end{center}
\vspace{-.5cm}
\caption{Dependence of the low-momentum in-medium NN interaction
on the nuclear density $\rho$. Shown are the momentum space matrix elements in
the $^1$S$_0$, $^3$S$_1$, and $^3$D$_1$ partial waves and the $^3$S$_1-^3$D$_1$
mixing matrix element.}
\label{sdwaves}
\end{figure}

In the $^3$D$_1$ partial wave we find again that the pion self-energy 
correction $V_{NN}^{\rm med, 1}$ largely cancels the effect from the long-range
Pauli-blocked vertex correction $V_{NN}^{\rm med, 2}$. The Pauli-blocked 
$2\pi$-exchange contribution $V_{NN}^{\rm med, 3}$ now vanishes at small 
momenta and has a large effect at high momenta, in contrast to its role
in the $S$-waves. By itself, $V_{NN}^{\rm med, 3}$ would nearly 
double the repulsion in the 
$^3$D$_1$ channel. This enhancement of the $^3$D$_1$ repulsion is reduced
in part by the attractive contribution coming from the medium-range 
Pauli-blocked one-pion-exchange vertex correction $V_{NN}^{\rm med, 4}$. The 
final possible contribution, $V_{NN}^{\rm med, 5}$, in this channel is
neglegible. Finally,
the $^3$S$_1-^3$D$_1$ mixing matrix element receives contributions from all 
density-dependent terms except the short-range contact interaction 
$V_{NN}^{\rm med, 6}$ since each of the first five contributions includes a
tensor component. Only $V_{NN}^{\rm med, 2}$ and $V_{NN}^{\rm med, 4}$ give 
repulsive contributions to the $^3$S$_1-^3$D$_1$ mixing, and these
terms are generally weaker than the three attractive contributions. 

\begin{figure}
\begin{center}
\includegraphics[height=12cm,angle=270]{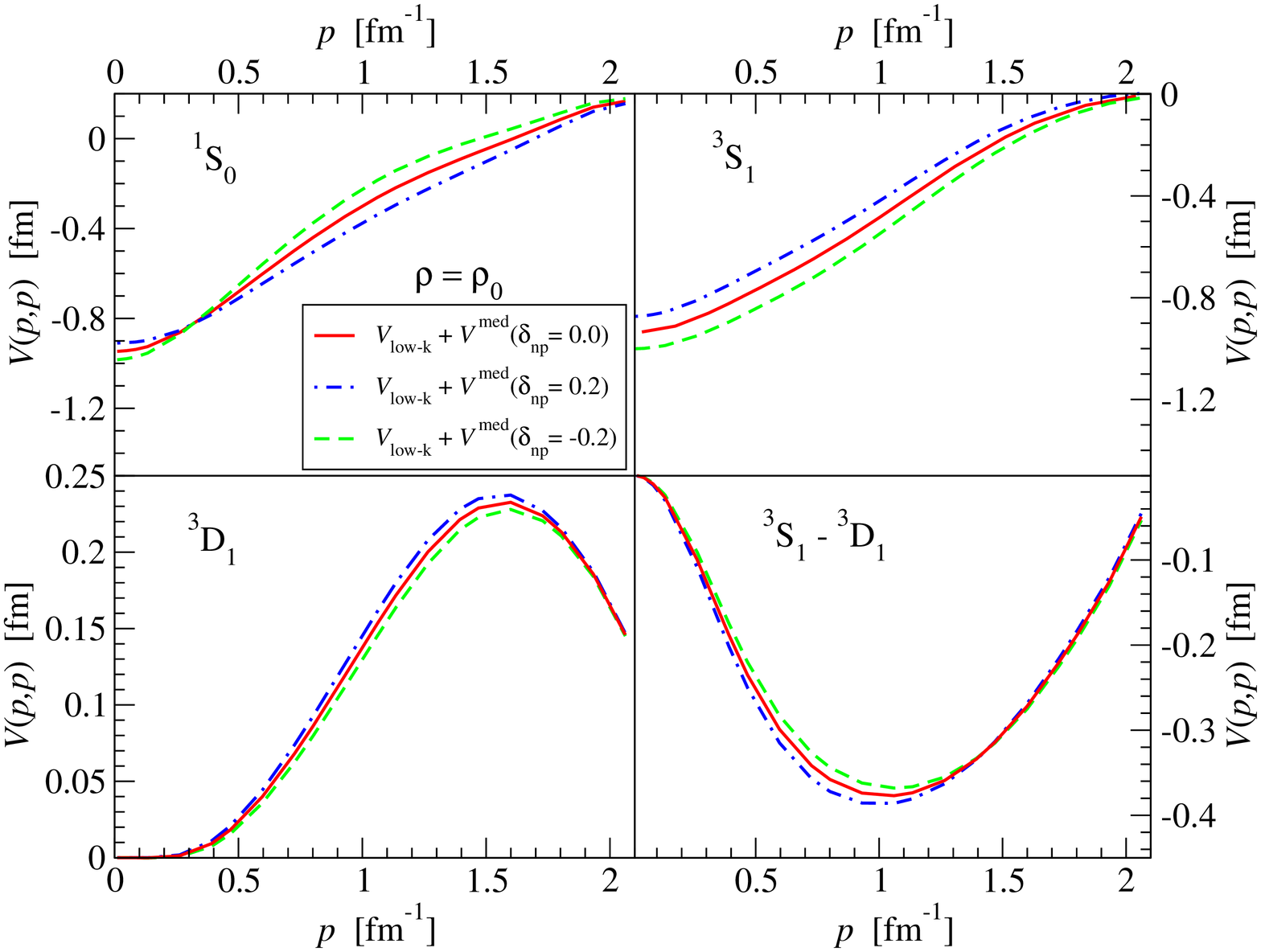}
\end{center}
\vspace{-.5cm}
\caption{Dependence of the low-momentum in-medium nucleon-nucleon interaction
on the isospin asymmetry $\delta_{np}=(\rho_n-\rho_p)/\rho$ at $\rho_0 
= 0.16$ fm$^{-3}$. Shown are the momentum space matrix elements in
the $^1$S$_0$, $^3$S$_1$, and $^3$D$_1$ partial waves and the $^3$S$_1-^3$D$_1$
mixing matrix element.}
\label{sdwavesi}
\end{figure}

In Fig.\ \ref{sdwaves} we show the modifications of the momentum-space 
matrix elements in the $^1$S$_0$, $^3$S$_1$, and $^3$D$_1$ partial waves
of \vlk due to all density-dependent 
contributions. The complete interaction is shown at the densities
$\rho_0/2$ and $\rho_0$. The overall effect is to decrease the
strong attraction in $S$-waves and to increase the repulsion in the (diagonal)
$^3$D$_1$ channel. The $^3$S$_1-^3$D$_1$ mixing matrix element becomes on average
only mildly more attractive. The repulsive effects increase with the nuclear
density and this way provide the mechanism for nuclear matter 
saturation when using the low-momentum interaction \vlk \cite{achim}. Finally, in
Fig.\ \ref{sdwavesi} we plot the in-medium $pp$
interaction at nuclear matter saturation density $\rho_0$ for isospin asymmetries
$\delta_{np} = \pm 0.2$, where $\delta_{np}=0.2$ corresponds to the asymmetry
reached in heavy nuclei. For larger isospin asymmetries, the effects scale linearly
with $\delta_{np}$. We plot as well the results for 
$\delta_{np} = -0.2$, which corresponds
to the $nn$ interaction in a medium with $\delta_{np}=0.2$.
As discussed in Section \ref{ianm} the effects on the $pp$ interaction
and $nn$ interaction are equal and of opposite sign. Although the effects
due to an isospin asymmetry are in general small, the difference between the 
$pp$ and $nn$ interactions can be significant, particularly in the $^3$S$_1$ channel.




\begin{figure}[t]
\begin{center}
\includegraphics[height=13cm,angle=270]{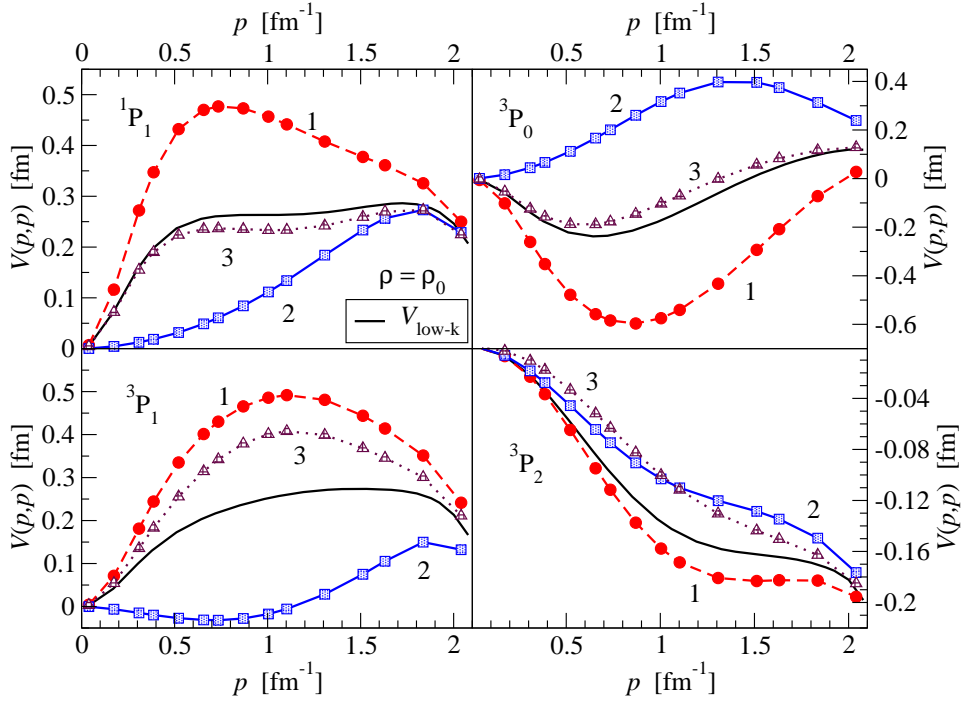}
\end{center}
\vspace{-.5cm}
\caption{Modifications to the $^1$P$_1$, $^3$P$_0$, $^3$P$_1$, and $^3$P$_2$ 
partial wave amplitudes of \vlk (denoted by the solid line) due to the
first three density-dependent
contributions $V_{NN}^{\rm med; \, 1,2,3}$ at a nuclear density $\rho=\rho_0$.}
\label{p123}
\end{figure}

\subsection{$P$-waves}
\begin{figure}
\begin{center}
\includegraphics[height=13cm,angle=270]{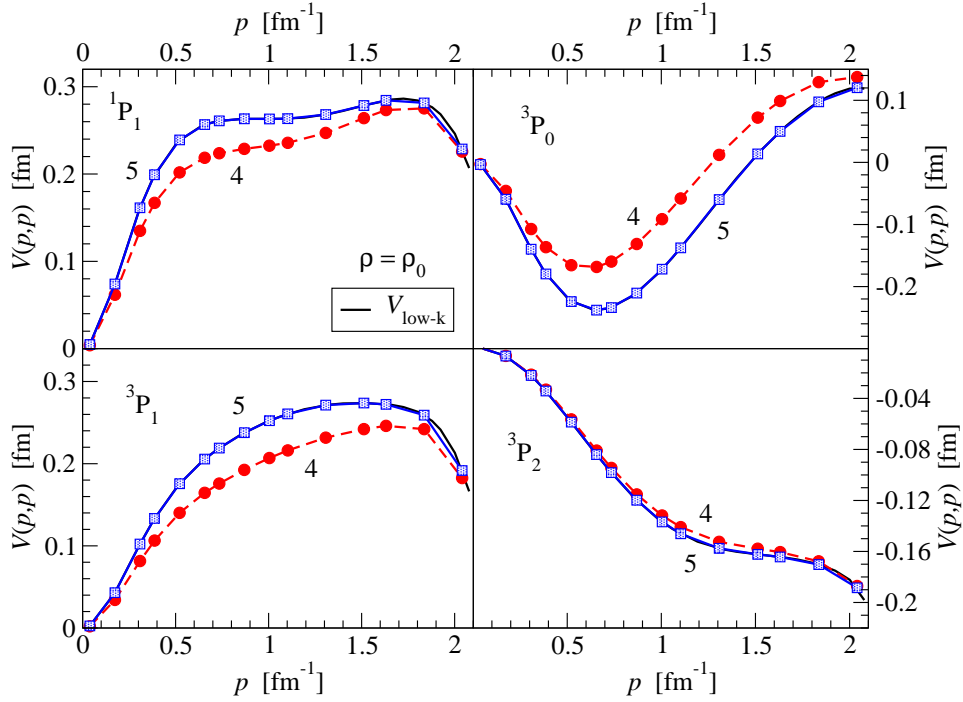}
\end{center}
\vspace{-.5cm}
\caption{Modifications to the $^1$P$_1$, $^3$P$_0$, $^3$P$_1$, and $^3$P$_2$ 
partial wave amplitudes of \vlk (shown by the solid line) due to 
$V_{NN}^{\rm med;\, 4,5}$ at a nuclear density $\rho=\rho_0$.}
\label{p456}
\end{figure}

There are three uncoupled $L=1$ partial waves, namely $^1$P$_1$, $^3$P$_0$, 
and $^3$P$_1$, as well as the $^3$P$_2$ partial wave which can couple through
the tensor and quadratic spin-orbit forces to the $^3$F$_2$ partial wave. In
Figs.\ \ref{p123} and \ref{p456} we have plotted the effects of the six
components of the density-dependent NN interaction on each of
these $P$-wave amplitudes. We consider a medium of isospin-symmetric nuclear matter
at saturation density $\rho_0$. Again we find that although the pion 
self-energy correction and the long-range one-pion-exchange vertex correction
are the largest contributions in all channels, taken together they have only a
moderate effect. The Pauli-blocked two-pion exchange contribution $V_{NN}^{\rm med, 3}$
is attractive in the $P$-wave spin-singlet states and repulsive in spin-triplet states,
though only in the $^3$P$_1$ and $^3$P$_2$ does it play a significant role.
Effects from the one-pion exchange vertex correction arising from the medium-range three-nucleon
force ($V_{NN}^{\rm med, 4}$) can be as large as 20\%, but those from the vertex-corrected
short-range nuclear force ($V_{NN}^{\rm med, 5}$) are negligible.

\begin{figure}
\begin{center}
\includegraphics[height=12cm,angle=270]{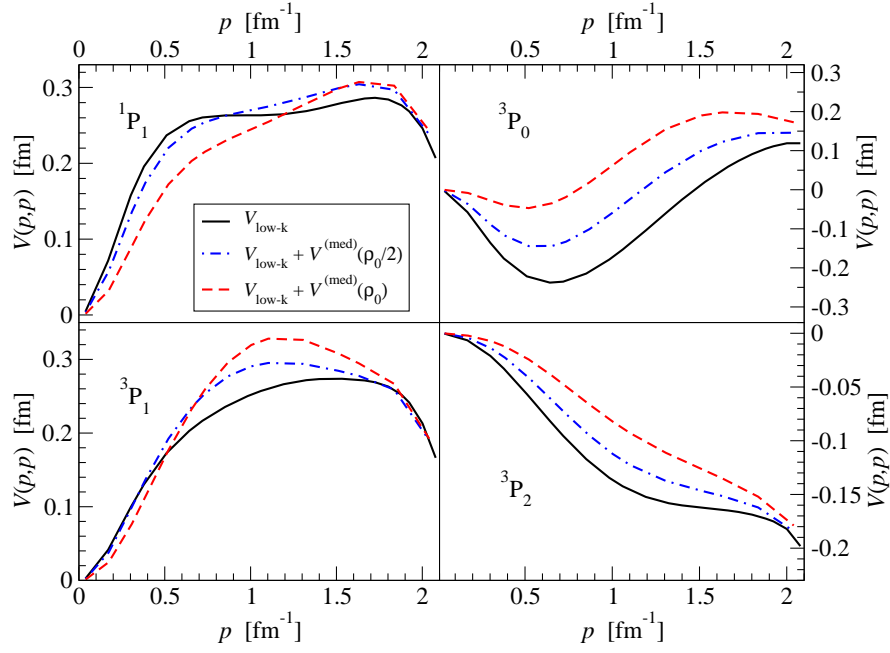}
\end{center}
\vspace{-.5cm}
\caption{Dependence of the low-momentum in-medium nucleon-nucleon interaction
on the nuclear density $\rho$. Shown are the momentum space matrix elements in
the $^1$P$_1$, $^3$P$_0$, $^3$P$_1$, and $^3$P$_2$ partial waves.}
\label{pwaves}
\end{figure}

\begin{figure}
\begin{center}
\includegraphics[height=12cm,angle=270]{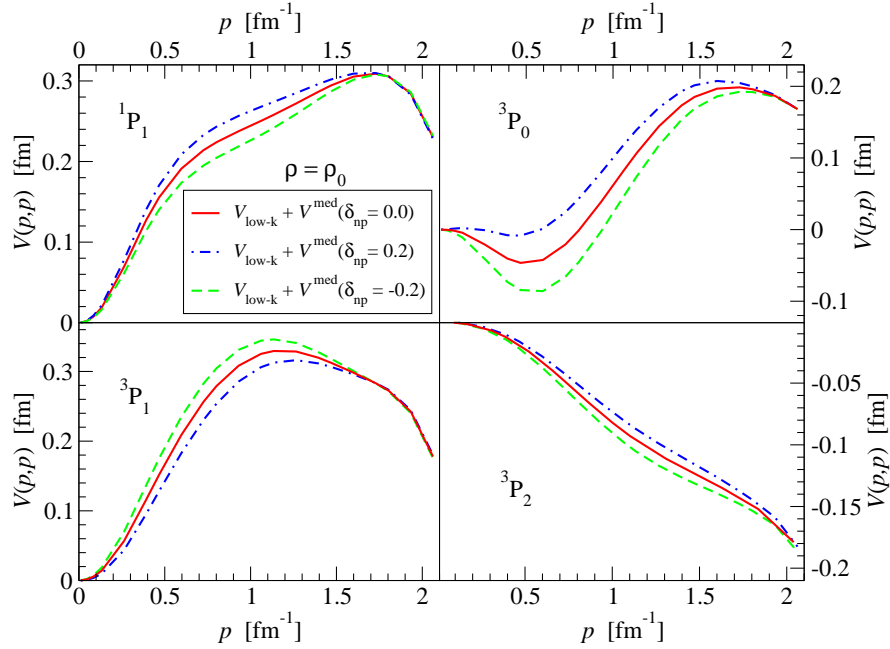}
\end{center}
\vspace{-.5cm}
\caption{Dependence of the low-momentum in-medium nucleon-nucleon interaction
on the isospin asymmetry $\delta_{np}=(\rho_n-\rho_p)/\rho$ at $\rho_0 
= 0.16$ fm$^{-3}$. Shown are the momentum space matrix elements in
the $^1$P$_1$, $^3$P$_0$, $^3$P$_1$, and $^3$P$_2$ partial waves.}
\label{pwavesi}
\end{figure}

In Fig.\ \ref{pwaves} we have plotted the complete $P$-wave interactions
at densities $\rho_0/2$ and $\rho_0$. Aside from the $^3$P$_1$ channel, which on average
receives a small attractive contribution from the sum of the six density-dependent
terms, we see
that the remaining $L=1$ partial waves all receive 
a net repulsive contribution. 
The $^3$P$_0$ 
partial wave is particularly sensitive to these modifications; at nuclear 
matter saturation density, nearly all of the attraction at small momenta vanishes
and the repulsion at higher momenta is largely increased. This results 
primarily from the repulsion due to $V_{NN}^{\rm med, 2}$, 
$V_{NN}^{\rm med, 3}$, and $V_{NN}^{\rm med, 4}$. The $^3$P$_1$ and
$^3$P$_2$ partial waves are less sensitive to the density dependent terms. 
The $^3$P$_1$ channel receives two attractive
($V_{NN}^{\rm med, 2}$, $V_{NN}^{\rm med, 4}$) and two repulsive 
($V_{NN}^{\rm med, 1}$, $V_{NN}^{\rm med, 3}$) modifications which give rise 
to only a small net repulsive effect limited to intermediate momenta. Finally, 
for the $^3$P$_2$ partial wave, only $V_{NN}^{\rm med, 1}$, 
$V_{NN}^{\rm med, 2}$, and $V_{NN}^{\rm med, 3}$ are important. The net 
repulsion from $V_{NN}^{\rm med, 2}$ and $V_{NN}^{\rm med, 3}$ is
approximately twice as large as the small attraction from 
$V_{NN}^{\rm med, 1}$. Isospin asymmetry
effects ($\delta_{np} = \pm 0.2$) in the $L=1$ partial waves are shown in
Fig.\ \ref{pwavesi}. The largest effect is in the $^3$P$_0$ channel where 
the difference between the $pp$ and $nn$ interaction is approximately 25\%
of the total interaction strength. In fact, in the $pp$ channel
there is almost no attraction remaining.


\subsection{$D$-waves and $D-G$ mixing}
\begin{figure}
\begin{center}
\includegraphics[height=13cm,angle=270]{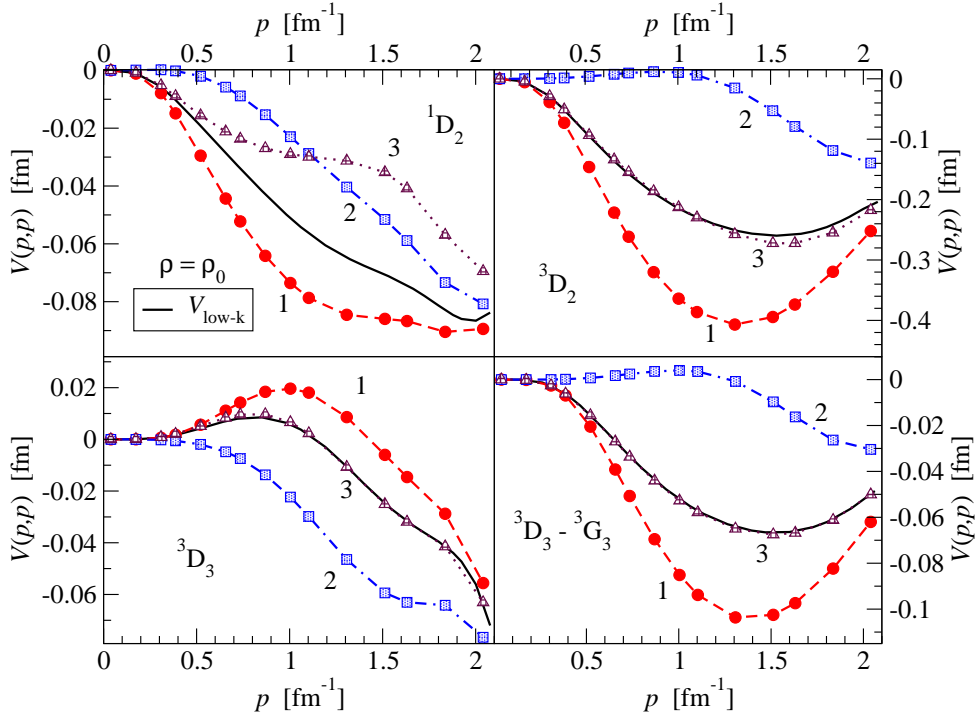}
\end{center}
\vspace{-.5cm}
\caption{Modifications to the $^1$D$_2$, $^3$D$_2$ and $^3$D$_3$ partial wave amplitudes and 
the $^3$D$_3-^3$G$_3$ mixing matrix element of \vlk (shown by the solid line) due to the 
first three density-dependent
contributions $V_{NN}^{\rm med;\, 1,2,3}$ at saturation density $\rho_0$.}
\label{d123}
\end{figure}

\begin{figure}
\begin{center}
\includegraphics[height=13cm,angle=270]{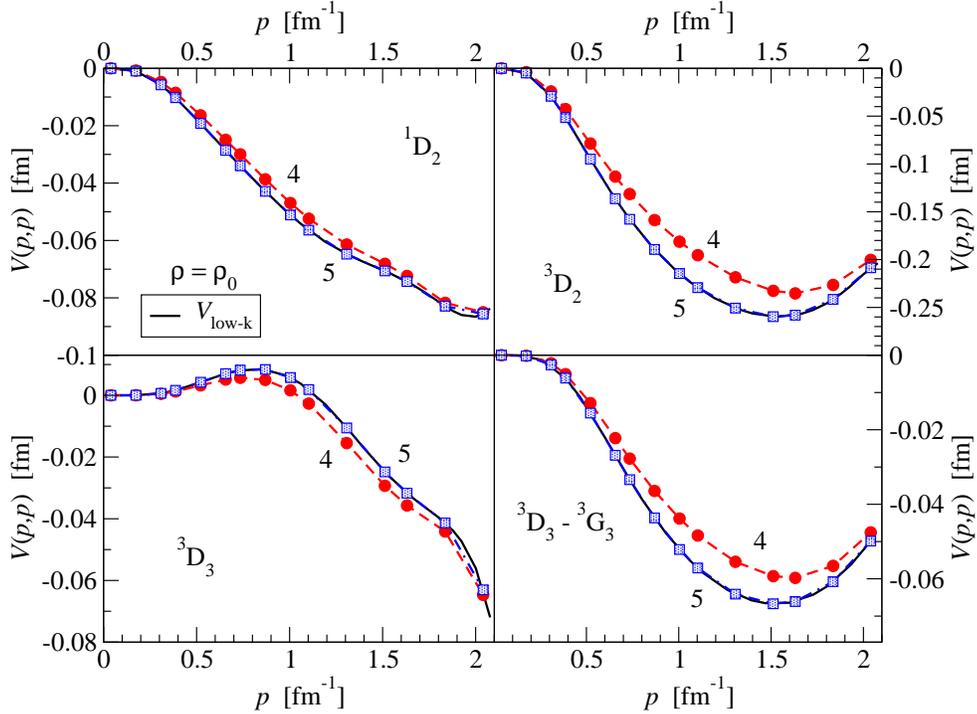}
\end{center}
\vspace{-.5cm}
\caption{Modifications to the $^1$D$_2$, $^3$D$_2$ and $^3$D$_3$ partial wave amplitudes and 
the $^3$D$_3-^3$G$_3$ mixing matrix element
of \vlk (shown by the solid line) due to 
$V_{NN}^{\rm med;\, 4,5}$ at saturation density $\rho_0$.}
\label{d456}
\end{figure}

\begin{figure}
\begin{center}
\includegraphics[height=12cm,angle=270]{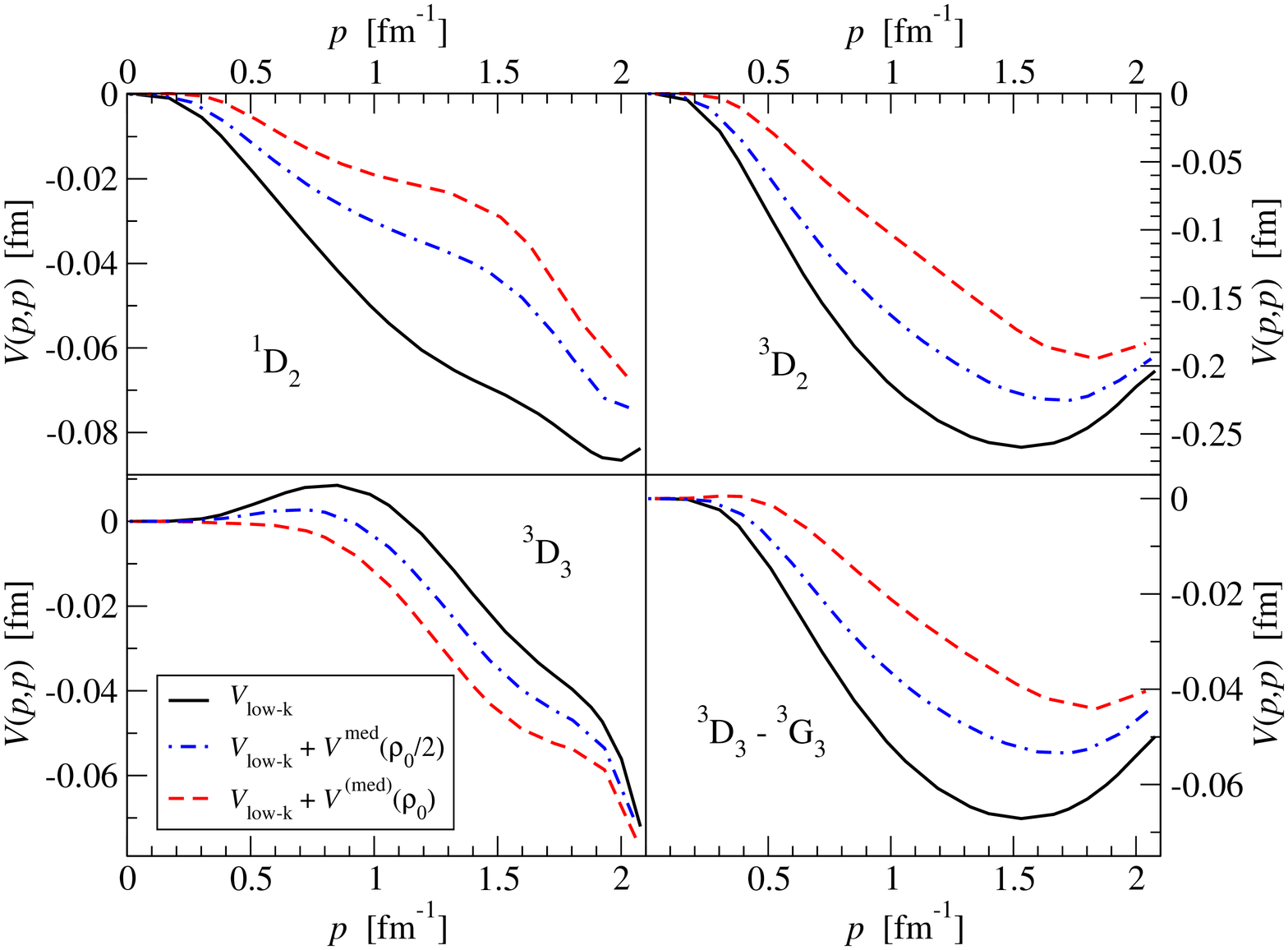}
\end{center}
\vspace{-.5cm}
\caption{Dependence of the low-momentum in-medium nucleon-nucleon interaction
on the nuclear density $\rho$. Shown are the momentum space matrix elements in
the $^1$D$_2$, $^3$D$_2$ and $^3$D$_3$ partial waves as well as the $^3$D$_3-^3$G$_3$
mixing matrix element.}
\label{dwaves}
\end{figure}

\begin{figure}
\begin{center}
\includegraphics[height=12cm,angle=270]{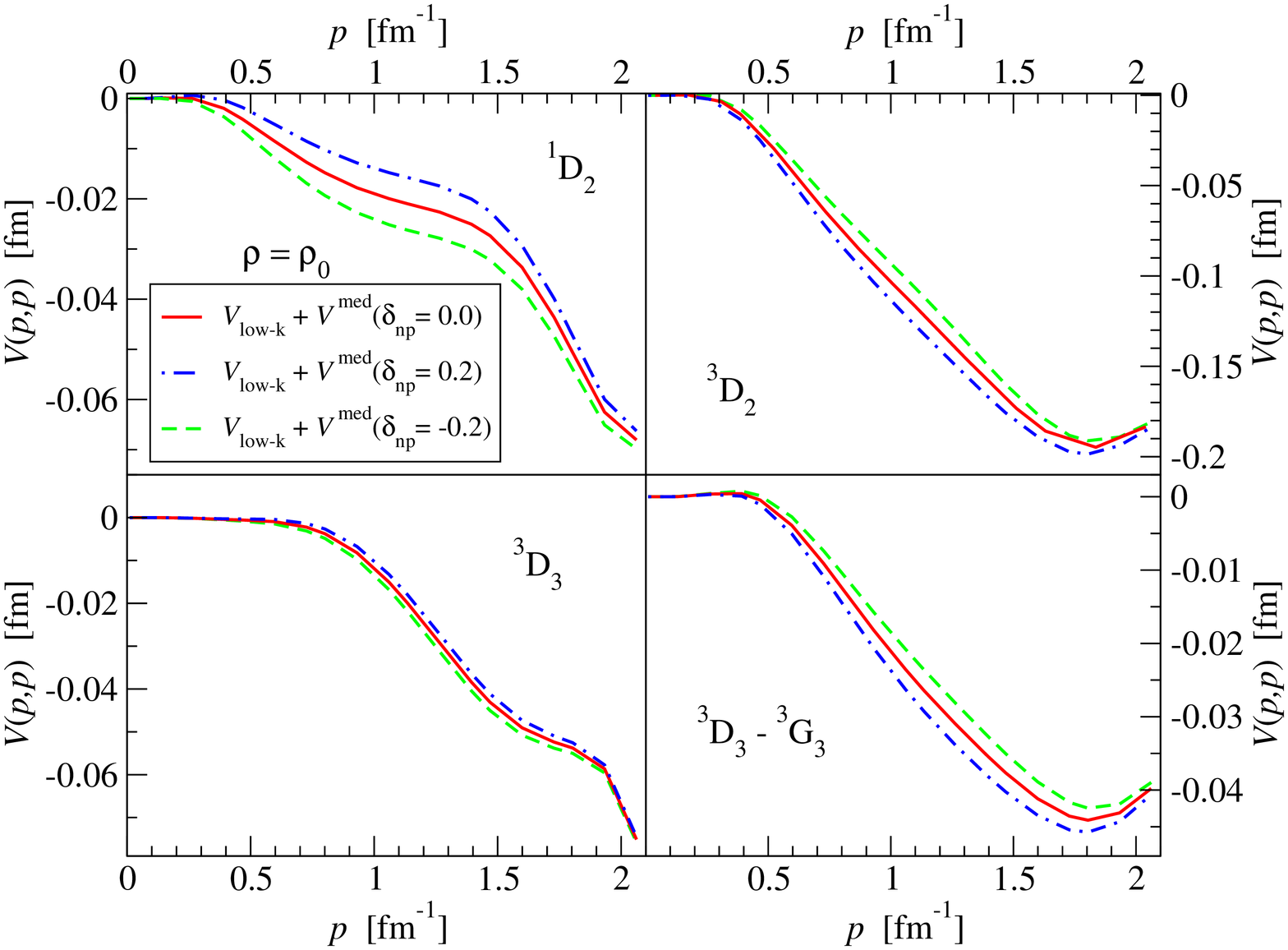}
\end{center}
\vspace{-.5cm}
\caption{Dependence of the low-momentum in-medium NN interaction
on the isospin asymmetry $\delta_{np}=(\rho_n-\rho_p)/\rho$ at $\rho_0 
= 0.16$ fm$^{-3}$. Shown are the momentum space matrix elements in
the $^1$D$_2$, $^3$D$_2$ and $^3$D$_3$ channels and the $^3$D$_3-^3$G$_3$ 
mixing matrix element.}
\label{dwavesi}
\end{figure}

Although the remaining $L=2$ partial waves are mostly smaller by an 
order of magnitude than the $L=0$ and 1 partial waves, it is 
important to see whether the trends observed in the lower partial waves 
continue. In Figs.\ \ref{d123} and \ref{d456} we have plotted 
the effects from the five (nonvanishing) density-dependent contributions in 
the $^1$D$_2$, $^3$D$_2$, and $^3$D$_3$ partial waves. As in the $S$ and $P$-waves, the 
$V_{NN}^{\rm med, 1}$ and $V_{NN}^{\rm med, 2}$ terms contribute strongest  
and with opposite sign. The long-range Pauli-blocked $2\pi$-exchange 
contribution $V_{NN}^{\rm med, 3}$ has little effect, except in the $^1$D$_2$
partial wave where it gives rise to a strong repulsion. Together with 
$V_{NN}^{\rm med, 2}$, the Pauli-blocked $2\pi$-exchange contribution gives
rise to a repulsion in this channel that is significantly larger than the 
attractive contribution from the pion self energy term $V_{NN}^{\rm med, 1}$.
In fact, both $V_{NN}^{\rm med, 2}$ and $V_{NN}^{\rm med, 4}$ are repulsive 
in all of these channels, except in the $^3$D$_3$ partial wave where they
combine to generate an overall attraction. These results can be seen 
from Fig.\ \ref{dwaves} in which we have plotted the
complete interaction at densities $\rho_0/2$ and $\rho_0$. From this figure 
one can see that all $L=2$ partial waves, except the $^3$D$_3$ channel, 
become less attractive due to the density-dependent corrections.
In Fig.\ \ref{dwavesi} we show the effect of an isospin asymmetry on these 
partial wave amplitudes for $\delta_{np}=\pm 0.2$. Again, we find that in
general the modifications are small, though in some channels, 
such as $^1$D$_2$, the difference between the $pp$ and $nn$ interactions
cannot be neglected.

\clearpage



\section{Summary and conclusions}
Using chiral effective field theory, we have derived in this work density-dependent
corrections to the nucleon-nucleon interaction in isospin-symmetric as well as isospin-asymmetric nuclear 
matter. These corrections have been calculated from the six one-loop in-medium
NN-scattering diagrams generated by the leading-order chiral three-nucleon interaction.
The resulting in-medium NN interaction has been transformed into the $|LSJ \rangle$ basis. 
Although we have combined the density-dependent corrections with the low-momentum NN potential
\vlk at a selected cutoff $\Lambda_{\rm low-k}=2.1$ fm$^{-1}$, the analytic expressions for the 
density-dependent terms should hold at any resolution scale by re-adjusting
the values of the running low-energy constants $c_D(\Lambda_{\rm low-k})$ and 
$c_E(\Lambda_{\rm low-k})$. In the present work the values of the two low-energy constants 
$c_D=-2.06$ and $c_E=-0.63$ were taken from a previous calculation \cite{nogga} of $A=3,4$ binding
energies using the low-momentum NN potential \vlk at $\Lambda_{\rm low-k}=2.1$ fm$^{-1}$.

After projecting the interaction into partial waves, we find that the largest 
density-dependent corrections come from the long-range 
Pauli-blocked pion self-energy and vertex correction. However, to a large extent they cancel 
in all partial waves studied and thus give rise to effects comparable to the 
remaining density-dependent corrections. The two contributions from the 
medium-range three-nucleon force proportional to $c_D$
are small and generally play only a minor role. In contrast to this, the long-range Pauli-blocked 
$2\pi$-exchange as well as the contact interaction proportional 
to $c_E$ provide significant repulsion in most partial waves. The
latter acts only in $S$-waves where it decreases the attraction
in the spin-singlet and spin-triplet channels by approximately 20\%. 
In fact, the net effect of all density-dependent corrections is repulsive in 
nearly all partial waves channels we studied. The
repulsive effects increase with density and in this way provide a mechanism 
for nuclear matter saturation \cite{achim}. The repulsive (stabilizing) nature of the pion-induced
three-body forces is also observed in the chiral perturbation theory calculation
of nuclear matter in ref.\ \cite{fritsch}.

Furthermore, we have found that additional medium modifications due 
to a (small) isospin asymmetry are present only in the $pp$ and $nn$ channels
and are generally small. However, the resulting difference between the 
in-medium $pp$ and $nn$ interactions can be significant in certain partial 
waves. This work should serve as a foundation for future nuclear structure
calculations exploiting density-dependent chiral effective two-nucleon interactions.

\end{document}